\documentclass[twocolumn,prb,aps,showpacs,amsmath,amssymb]{revtex4}

\usepackage{graphicx}
\usepackage{dcolumn}
\usepackage{bm}

\begin{document}

\title{Slave-boson Keldysh field theory for the Kondo effect in quantum dots}

\author{Sergey Smirnov}
\email{sergey.smirnov@physik.uni-regensburg.de}
\author{Milena Grifoni}
\affiliation{Institut f\"ur Theoretische Physik, Universit\"at Regensburg,
  D-93040 Regensburg, Germany}

\date{\today}

\begin{abstract}
We present a {\it nonequilibrium nonperturbative} field theory for the Kondo
effect in strongly interacting quantum dots at finite temperatures. Unifying
the slave-boson representation with the Keldysh field integral an effective
Keldysh action is derived and explored in the vicinity of the zero
slave-bosonic field configuration. The theory properly reflects the essential
features of the Kondo physics and at the same time significantly simplifies a
field-theoretic  treatment of the phenomenon, avoiding complicated saddle
point analysis or $1/N$ expansions, used so far. Importantly, our theory
admits a {\it closed analytical} solution which explains the mechanism of the
Kondo effect in terms of an interplay between the real and imaginary parts
of the slave-bosonic self-energy. It thus provides a convenient nonperturbative
building block, playing the role of a "free propagator", for more advanced
theories. We finally demonstrate that already this simplest possible field
theory is able to correctly reproduce experimental data on the Kondo peak
observed in the differential conductance, correctly predicts the Kondo
temperature and, within its applicability range, has the same universal
temperature dependence of the conductance as the one obtained in numerical
renormalization group calculations.
\end{abstract}

\pacs{72.15.Qm, 73.63.-b, 72.10.Fk}

\maketitle

\section{Introduction}\label{intro}
Experimentally discovered almost eighty years ago \cite{de_Haas_1934} and
qualitatively explained thirty years after \cite{Kondo_1964}, the Kondo
effect, a minimum in the temperature dependence of the resistance in magnetic
alloys, was later anew brought into the world and garbed with physics of
quantum dots (QD) \cite{Ralph_1994,Goldhaber-Gordon_1998}. That time this
complicated phenomenon was supplemented by physics of nonequilibrium due to
the dot coupling to contacts biased by an external voltage. In this setup the
Kondo effect, a nonperturbative phenomenon induced by both the
electron-electron interactions and the QD-contacts coupling, appears as a
sharp many-particle resonance in the tunneling density of states (TDOS) at the
Fermi energy. An immediate consequence of this resonance is the zero-bias
maximum observed experimentally in the QD differential conductance at low
temperatures.

Theoretical predictions \cite{Glazman_1988,Meir_1993,Wingreen_1994} of this
behavior made before the actual experiments use the single-impurity Anderson
model (SIAM) \cite{Anderson_1961} with the on-dot interaction $U$. Those early
works utilized either a quasi-particle transformation \cite{Glazman_1988} to
analytically predict resonant transmission through QDs for arbitrary $U$ or
the noncrossing approximation (NCA), equations of motion and perturbation
theory \cite{Meir_1993,Wingreen_1994,Hettler_1995,Sivan_1996,Entin_2005} to
numerically and analytically describe several important stationary and
nonstationary features of the Kondo effect in strongly interacting QDs modeled
by the infinite-$U$ SIAM, in particular, in a slave-boson representation
\cite{Coleman_1984,Coleman_1987,Hewson_1997}.

Later, as an alternative to the slave-boson representation, diagrammatic
expansions within the reduced density matrix formalism \cite{Koenig_1998} were
applied to SIAM for the infinite-$U$ case. Extracting a certain infinite
subset of diagrams, an analytical nonperturbative expression for the TDOS was
obtained at finite temperatures. Here the absence of the double occupancy
played a crucial role in identifying the relevant diagrams.
\begin{figure}
\includegraphics[width=7.6 cm]{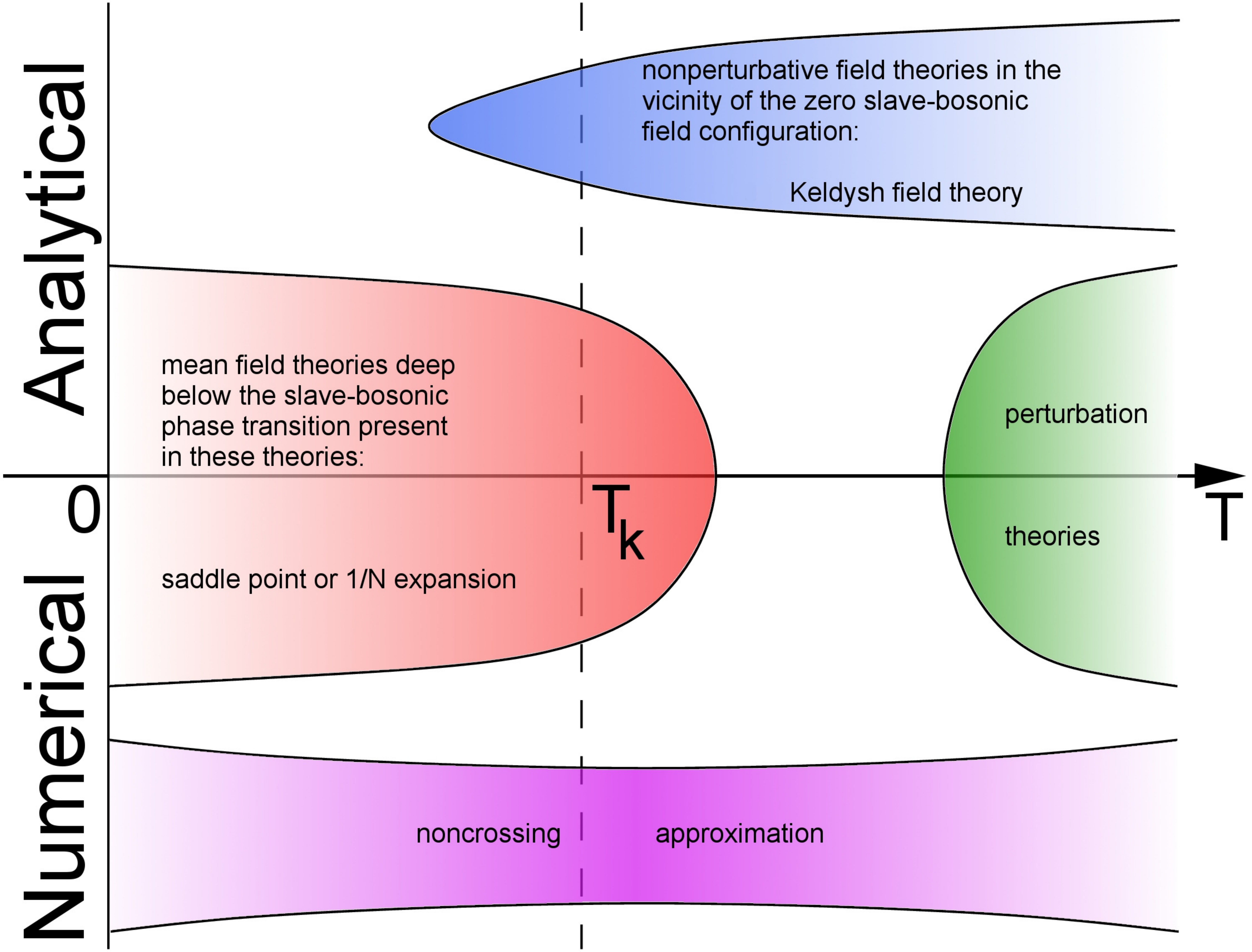}
\caption{\label{figure_1} (Color online) Classes of the slave-bosonic theories
  for the Kondo effect in QDs. Their range of quantitative reliability with
  respect to the Kondo temperature $T_\text{K}$
  ($T_\text{K}/\Gamma\thicksim\exp[-2\pi(\mu_0-\epsilon_\text{d})/\Gamma]$,
  where $\Gamma$ is the total QD-contacts coupling strength, which is twice
  that of Ref. \onlinecite{Meir_1993}, $\mu_0$ the QD chemical potential,
  $\epsilon_\text{d}$ the single-particle energy level of the QD) is given by
  their horizontal location. The vertical location of a class shows its type,
  numerical, analytical or both. As to our field theory, what the blue class
  shows is what one generally expects from nonperturbative field theories in
  the vicinity of the zero slave-bosonic field configuration.}
\end{figure}

For small, intermediate and large $U$ the Kondo problem in QDs was widely
explored numerically
\cite{Gezzi_2007,Anders_2008,Heidrich_2009,Eckel_2010,Muehlbacher_2011}.

In the present paper we develop an analytical nonequilibrium real-time field
theory of the Kondo effect in strongly interacting QDs at finite temperatures
using the infinite-$U$ SIAM. The field-theoretic approach is based on the
slave-boson representation and the Keldysh field integral
\cite{Kamenev_1999,Altland_2010}.

Importantly, our theory is nonperturbative in both the electron-electron
interaction and QD-contacts coupling. The necessity in such a theory is
obvious from its advantages. Firstly, it is a field theory and, thus, it has a
clear systematic generalization from the present relatively simple basic model
to more involved setups such as the ones with ferromagnetic contacts,
superconducting contacts, finite $U$ systems, etc. This is especially
important since there is a limited number of analytical theories
nonperturbative in both the electron-electron interaction and QD-contacts
coupling and which at the same time have a straightforward generalization
scheme. Secondly, it is an analytical theory and, thus, it will help to reveal
a relevant physical picture behind new and more complicated physical systems
as, {\it e.g.}, those mentioned above. This is definitely an advantage over
numerical methods such as, {\it e.g.}, NCA, which provide good quantitative
description but hide the essence of the physics, giving only an indirect
access to it.

In Fig. \ref{figure_1} we show a comparative layout of the slave-bosonic
theories for the Kondo effect in QDs. A basic nonperturbative analytical
slave-bosonic field theory of the Kondo effect in QDs within the range from
below $T_\text{K}$ to higher temperatures, {\it i.e.}, within the most
relevant experimental temperature range (blue (uppermost) class in Fig.
\ref{figure_1}), is formulated in this work.

So far, using the slave-boson method in the context of QDs, the results were
obtained nonperturbatively numerically
\cite{Meir_1993,Wingreen_1994,Hettler_1995,Aguado_2000} below and/or above
$T_\text{K}$, perturbatively semi-analytically \cite{Sivan_1996} above
$T_\text{K}$ and nonperturbatively analytically \cite{Ratiani_2009} below
$T_\text{K}$.

The slave-boson approach excludes the double occupancy of a QD restricting its
Hilbert space to zero and single occupancy. This restriction was taken into
account exactly in the numerical solutions
\cite{Meir_1993,Wingreen_1994,Hettler_1995} using an additional integral
removing the constraint \cite{Bickers_1987}. However, since the mean field
\cite{Aguado_2000} and 1/$N$ expansion \cite{Ratiani_2009}
theories, using the same integral trick, impose the constraint only
approximately, they fail at higher temperatures \cite{Hewson_1997}.

Here our goal is the development of an analytical ground for the slave-bosonic
nonperturbative field theories which could be placed within the blue (uppermost)
class in Fig. \ref{figure_1}. To avoid problems with high temperatures we do not
use the integral trick to take into account the restriction of the Hilbert space
but instead use an alternative method based on taking a certain limit with
respect to a real parameter \cite{Bickers_1987}. This alternative method has
not been applied so far in conjunction with the Keldysh field integral. We
demonstrate that the corresponding limit can be exactly performed analytically
after the Keldysh field integral for the TDOS has been analytically calculated
using a certain approximation.

Specifically, the approximation concerns the effective Keldysh action obtained
after integrating out all electronic degrees of freedom. This action being a
nonlinear functional of the slave-bosonic field is expanded up to second order
in this field. Note, that this does not imply any perturbation since the
action itself is the argument of an exponent. According to the general concept
of the condensed matter field theory \cite{Altland_2010} the physics of our
model is the physics in the vicinity of the zero slave-bosonic field
configuration. We demonstrate that this physics contains the Kondo effect
in QDs at finite temperatures. This scenario complements \cite{Altland_2010}
the saddle point analysis \cite{Ratiani_2009} valid deep below the
mean field theory slave-bosonic phase transition, i.e., at low temperatures
when the slave-bosonic field in the mean field theory is
condensed. Surprisingly, our simple or "bare" theory, which might play a role
of a nonperturbative "free propagator" for more advanced nonperturbative
theories of the blue class in Fig. \ref{figure_1}, already provides a good
description of the Kondo peak observed \cite{Ralph_1994} in the
differential conductance at temperatures close to $T_\text{K}$.

The paper is organized as follows. In Section \ref{of} we formulate the
problem on the operator level. Then in Section \ref{fts} we translate this
formulation into a field-theoretic framework using the Keldysh field
integral and derive an analytic expression for the TDOS. Finally, we discuss
our results and make conclusions in Sections \ref{dr} and \ref{concl},
respectively.

\section{Operator formulation: Hamiltonian and observables}\label{of}
We start with the infinite-$U$ Anderson Hamiltonian in a slave-boson
representation. As is well known \cite{Coleman_1984}, in the case when
$U=\infty$, the Anderson Hamiltonian,
$\hat{H}_\text{d}=\sum_\sigma\epsilon_\text{d}\hat{n}_{\text{d},\sigma}+U\hat{n}_{\text{d},\uparrow}\hat{n}_{\text{d},\downarrow}$,
where $\hat{n}_{\text{d},\sigma}=d^\dagger_\sigma d_\sigma$,
$\sigma=\uparrow,\downarrow$, takes the form
$\hat{H}_\text{d}=\sum_\sigma\epsilon_\text{d}f^\dagger_\sigma f_\sigma$,
where the new fermionic operators are related to the original ones as
$d_\sigma=f_\sigma b^\dagger$, $d^\dagger_\sigma=f^\dagger_\sigma b$, and $b$,
$b^\dagger$ are the annihilation and creation operators of a
slave-boson. Using these new fermionic and slave-bosonic operators the
tunneling Hamiltonian, describing the coupling of the infinite-$U$ Anderson QD
to contacts, can be written as
\begin{equation}
\hat{H}_\text{T}=\sum_{a\sigma}\bigl(T_{a\sigma}c^\dagger_af_\sigma b^\dagger+T^*_{a\sigma}f^\dagger_\sigma c_a b\bigl),
\label{tun_Ham}
\end{equation}
where $c_a$, $c_a^\dagger$ are the annihilation and creation operators of the
contacts fermions, $a$ includes the contacts complete set of quantum numbers
and the contacts labels, left (L) or right (R), and $T_{a\sigma}$ are the
tunneling matrix elements. The contacts are described by the Hamiltonian
$\hat{H}_\text{C}=\sum_a\epsilon_ac^\dagger_ac_a$ and are assumed to be in
equilibrium with the chemical potentials $\mu_\text{L,R}=\mu_0-eV_\text{L,R}$,
with $V\equiv V_\text{L}-V_\text{R}$ being an external voltage. The total
Hamiltonian is $\hat{H}=\hat{H}_\text{d}+\hat{H}_\text{T}+\hat{H}_\text{C}$.

The restriction of the QD Hilbert space to zero and single occupancy requires
the total number of the new fermions and slave-bosons, $\hat{Q}\equiv
b^\dagger b+\sum_\sigma f^\dagger_\sigma f_\sigma$, to be equal to one,
$\hat{Q}=\hat{I}$. This restriction must be taken into account in a QD
observable $\langle\hat{O}\rangle$. There are two ways of doing that
\cite{Bickers_1987}. In the QD context only the first one, {\it i.e.}, the
integral way was used so far
\cite{Meir_1993,Wingreen_1994,Hettler_1995,Aguado_2000,Ratiani_2009}. Here we
employ the second way from Ref. \onlinecite{Bickers_1987},
\begin{equation}
\langle\hat{O}\rangle(t)=
\frac{\underset{\mu\rightarrow\infty}{\text{lim}}
e^{\beta\mu}\text{Tr}[\hat{U}_{-\infty,t}\hat{O}\hat{U}_{t,-\infty}\hat{\rho}_0e^{-\beta\mu\hat{Q}}]}
{\underset{\mu\rightarrow\infty}{\text{lim}}e^{\beta\mu}\text{Tr}[\hat{\rho}_0\hat{Q}e^{-\beta\mu\hat{Q}}]},
\label{observ_hsr}
\end{equation}
where $\hat{U}_{t,t'}$, is the evolution operator with respect to the
Hamiltonian $\hat{H}$,
$\hat{\rho}_0=\exp[-\beta(\hat{H}_\text{d}-\mu_0\hat{N}_\text{d})]\otimes\exp[-\beta(\hat{H}_\text{C}-\sum_x\mu_x\hat{N}_x)]$,
($x=\text{L},\text{R}$) is the initial statistical operator with
$\hat{N}_\text{d}$ and $\hat{N}_x$ being the number operators of the QD and
contacts and $\beta$ is the inverse temperature. Eq. (\ref{observ_hsr}) may be
rewritten in two equivalent forms,
\begin{equation}
\langle\hat{O}\rangle(t)=
\frac{\underset{\mu\rightarrow\infty}{\text{lim}}
e^{\beta\mu}\text{Tr}[\hat{U}_{-\infty,\infty}\hat{U}_{\infty,t}\hat{O}\hat{U}_{t,-\infty}\hat{\rho}_0e^{-\beta\mu\hat{Q}}]}
{\underset{\mu\rightarrow\infty}{\text{lim}}e^{\beta\mu}\text{Tr}[\hat{\rho}_0\hat{Q}e^{-\beta\mu\hat{Q}}]},
\label{observ_hsr_fb}
\end{equation}
and
\begin{equation}
\langle\hat{O}\rangle(t)=
\frac{\underset{\mu\rightarrow\infty}{\text{lim}}
e^{\beta\mu}\text{Tr}[\hat{U}_{-\infty,t}\hat{O}\hat{U}_{t,\infty}\hat{U}_{\infty,-\infty}\hat{\rho}_0e^{-\beta\mu\hat{Q}}]}
{\underset{\mu\rightarrow\infty}{\text{lim}}e^{\beta\mu}\text{Tr}[\hat{\rho}_0\hat{Q}e^{-\beta\mu\hat{Q}}]}.
\label{observ_hsr_bb}
\end{equation}
These two forms have different interpretation: in Eq. (\ref{observ_hsr_fb})
the observable is taken within the evolution from $-\infty$ to $\infty$ while
in Eq. (\ref{observ_hsr_bb}) it is taken within the evolution from $\infty$ to
$-\infty$.

To develop a Keldysh field theory one first equivalently rewrites
Eqs. (\ref{observ_hsr_fb}) and (\ref{observ_hsr_bb}) on the Keldysh contour
$C_\text{K}$. To this end one gives the creation and annihilation operators a
formal temporal argument to allow the time-ordering operator to appropriately
interlace operators. Afterwards one may take the half-sum of the two
equivalent expressions to get the following symmetric form:
\begin{equation}
\begin{split}
\langle\hat{O}\rangle(t)=\frac{1}{\mathcal{N}_0}\underset{\mu\rightarrow\infty}{\text{lim}}
&\frac{e^{\beta\mu}}{\text{Tr}[\hat{\rho}'_0(\mu)]}
\text{Tr}[T_{C_\text{K}}e^{-\frac{i}{\hbar}\int_{C_\text{K}}d\tau\hat{H}'(\tau)}\times\\
&\times\frac{\hat{O}(t_+)+\hat{O}(t_-)}{2}\hat{\rho}'_0(\mu)],
\end{split}
\label{observ_hsr_1}
\end{equation}
where $t_+$ and $t_-$ are the projections of $t$ onto the forward and backward
branches of $C_\text{K}$,
\begin{equation}
\begin{split}
&\hat{\rho}'_0(\mu)=\hat{\rho}_0\exp(-\beta\mu\hat{Q}),\quad\hat{H}'=\hat{H}+\mu\hat{Q},\\
&\frac{1}{\mathcal{N}_0}\equiv\frac{\text{Tr}(\hat{\rho}_\text{C})}{\underset{\mu\rightarrow\infty}{\text{lim}}\{\exp(\beta\mu)\text{Tr}[\hat{Q}\hat{\rho}'_0(\mu)]\}}
\end{split}
\label{so_ham_pr}
\end{equation}
and $\hat{\rho}_\text{C}$ is the statistical operator of the contacts. The
expression under the limit in Eq. (\ref{observ_hsr_1}) can be written as the
Keldysh field integral \cite{Kamenev_1999,Altland_2010} for a fixed value of
$\mu$. The basic steps in the construction of the Keldysh field integral are
identical to the ones presented in Ref. \onlinecite{Altland_2010}. The details
specific to our application of this field integral are given in the next
section.

We would like to emphasize the absence in Eq. (\ref{observ_hsr_1}) of any
prefactor depending on the QD-contacts coupling. This is a great advantage of
the Keldysh field theory over non-field-theoretic approaches
\cite{Wingreen_1994} and imaginary-time field theories
\cite{Bickers_1987}. Indeed, in our approach there is no need for an
independent calculation of such prefactors. This greatly simplifies the
analytical exact projection onto the physical subspace. This fact was not
realized in the first attempt \cite{Ratiani_2009} to combine the Keldysh field
theory and slave-boson approach and, as a result, this attempt was reduced to
a nonequilibrium analog of equilibrium imaginary-time field theories with only
an approximate projection onto the physical subspace.

\section{Field-theoretic solution: Keldysh field integral}\label{fts}
We obtain the effective Keldysh field theory in a way similar to the one used
in Refs. \onlinecite{Altland_2010,Altland_2009} for Coulomb-blockaded
QDs. Namely, we first integrate out the QD and contacts Grassmann
fields. After this step the field-theoretic description is given in terms of
the effective Keldysh action,
$S_\text{eff}[\chi^\text{cl}(t),\chi^\text{q}(t)]\equiv
S_\text{B}^{(0)}[\chi^\text{cl}(t),\chi^\text{q}(t)]+S_\text{tun}[\chi^\text{cl}(t),\chi^\text{q}(t)]$,
where $\chi^{\text{cl},\text{q}}(t)$ are the classical and quantum components
of the slave-bosonic complex field, which is just a bosonic coherent state
\cite{Altland_2010}, $S_\text{B}^{(0)}[\chi^\text{cl}(t),\chi^\text{q}(t)]$ is
the free slave-bosonic action with the standard matrix form in the Keldysh
space and $S_\text{tun}[\chi^\text{cl}(t),\chi^\text{q}(t)]$ is the slave-bosonic
tunneling action,
\begin{equation}
S_\text{tun}[\chi^\text{cl}(t),\chi^\text{q}(t)]=-i\hbar\,\text{tr}\ln\bigl[I+\mathcal{T}G^{(0)}\bigl],
\label{sbta_ex}
\end{equation}
where the trace and matrix product are taken with respect to the temporal
arguments and both single-particle and Keldysh indices. In Eq. (\ref{sbta_ex})
the matrices $G^{(0)}$ and $\mathcal{T}$ have the block form in the
dot-contacts space,
\begin{equation}
G^{(0)}=
\begin{pmatrix}
G^{(0)}_\text{d}(\sigma t|\sigma't')&0\\
0&G^{(0)}_\text{C}(at|a't')
\end{pmatrix},
\label{G0_matr}
\end{equation}
\begin{equation}
\mathcal{T}=
\begin{pmatrix}
0&M_\text{T}^\dagger(\sigma t|a't')\\
M_\text{T}(at|\sigma't')&0
\end{pmatrix},
\label{T_matr}
\end{equation}
where the blocks $G^{(0)}_\text{d,C}(\alpha t|\alpha't')$ are the standard
fermionic Keldysh Green's function matrices ($2\times 2$ matrices in the
Keldysh space) of the free QD ($\alpha=\sigma$) and contacts ($\alpha=a$),
\begin{equation}
G^{(0)}_\text{d,C}(\alpha t|\alpha't')\!=\!
\begin{pmatrix}
G^{(0)+}_\text{d,C}(\alpha t|\alpha't')&G^{(0)\text{K}}_\text{d,C}(\alpha t|\alpha't')\\
0&G^{(0)-}_\text{d,C}(\alpha t|\alpha't')
\end{pmatrix},
\label{f_KGf_matr}
\end{equation}
with $G^{(0)+,-,\text{K}}_\text{d,C}(\alpha t|\alpha't')$ being the retarded,
advanced and Keldysh components \cite{Altland_2010}, respectively, and the
block $M_\text{T}(at|\sigma t')$ is the tunneling matrix ($2\times 2$ matrix
in the Keldysh space),
\begin{equation}
M_\text{T}(at|\sigma t')=\frac{1}{\hbar}\delta(t-t')T_{a\sigma}\frac{1}{\sqrt{2}}
\begin{pmatrix}
\bar{\chi}^\text{cl}(t)&\bar{\chi}^\text{q}(t)\\
\bar{\chi}^\text{q}(t)&\bar{\chi}^\text{cl}(t)
\end{pmatrix}.
\label{tun_matr}
\end{equation}
Note that the only formal difference of Eq. (\ref{tun_matr}) from the
corresponding expression in Ref. \onlinecite{Altland_2010} is that here
instead of the bosonic phase field we have the slave-bosonic field and
that this field is not exponentiated.

Our goal is to investigate a strongly interacting QD in the Kondo regime when
the Kondo resonance is not strong yet. This is the case, {\it e.g.}, at
temperatures above the Kondo temperature when the Kondo resonance is already
present but not fully developed. These range of temperatures is most relevant
for both experiments and practical applications in devices. As the formation
of the Kondo resonance is related to the QD population oscillations induced by
the QD-contacts coupling, the QD empty state will weakly fluctuate and its
probability will be small in the regime which we are interested in. Since the
QD empty state is described by the slave-bosonic complex field, those weak
oscillations, induced by the QD-contacts tunneling coupling, can be obtained
from the expansion of the tunneling action around the zero slave-bosonic field
configuration. This expansion is valid for small amplitudes of the
slave-bosonic complex field and thus for small probabilities of the QD empty
state.

We then expand the tunneling action (\ref{sbta_ex}) around the zero
slave-bosonic field configuration, {\it i.e.}, we keep in this action only the
first non-vanishing term which is quadratic in the slave-bosonic
field. Additionally, we assume a Lorentzian contacts density of states,
$\nu_\text{C}(\epsilon)=\nu_\text{C}D(\epsilon)$,
$D(\epsilon)=W^2/(\epsilon^2+W^2)$, $a=\{x,k,\sigma\}$ and
$T_{xk\sigma\sigma'}=\delta_{\sigma\sigma'}T$. The expression for the
tunneling matrix elements, in particular, means that we, for simplicity,
consider the case of a symmetric coupling to the contacts. The tunneling
action then becomes
\begin{equation}
\begin{split}
&S_\text{tun}[\chi^\text{cl}(t),\chi^\text{q}(t)]=(g/\hbar)\!\!\int \!\!dt\!\!\int \!\!dt'
\begin{pmatrix}
\bar{\chi}^\text{cl}(t)&\bar{\chi}^\text{q}(t)
\end{pmatrix}\times\\
&\times\begin{pmatrix}
0&\Sigma^-(t-t')\\
\Sigma^+(t-t')&\Sigma^\text{K}(t-t')
\end{pmatrix}
\begin{pmatrix}
\chi^\text{cl}(t')\\
\chi^\text{q}(t')
\end{pmatrix}.
\end{split}
\label{sbta}
\end{equation}
In Eq. (\ref{sbta}) $g\equiv 2\pi^2\nu_\text{C}|T|^2$ and the retarded,
advanced and Keldysh slave-bosonic self-energies are
\begin{equation}
\begin{split}
&\Sigma^\pm(t-t')\equiv(i/2)\sum_x[g^\text{K}_x(t'-t)g^\pm_\text{d}(t-t')+\\
&+g^\mp(t'-t)g^\text{K}_\text{d}(t-t')],\\
&\Sigma^\text{K}(t-t')\equiv(i/2)\sum_x\{g^\text{K}_x(t'-t)g^\text{K}_\text{d}(t-t')-\\
&\!\!\!-[g^+_\text{d}(t-t')-g^-_\text{d}(t-t')][g^+(t'-t)-g^-(t'-t)]\},
\end{split}
\label{sb_se_def}
\end{equation}
where the retarded, advanced and Keldysh components of the Green's function
matrix are
\begin{equation}
\begin{split}
&g^\pm(\omega)\equiv\int\frac{d\epsilon}{2\pi}D(\epsilon)\frac{\hbar}{\hbar\omega-\epsilon\pm i0},\\
&g^\pm_\text{d}(\omega)\equiv\frac{\hbar}{\pi(\hbar\omega-\epsilon_\text{d}-\mu\pm i0)},
\end{split}
\label{gf_ra}
\end{equation}
\begin{equation}
\begin{split}
&g^\text{K}_{x}(\omega)\equiv[g^+(\omega)-g^-(\omega)]\tanh\biggl[\frac{\beta(\hbar\omega-\mu_x)}{2}\biggl],\\
&g^\text{K}_\text{d}(\omega)\equiv[g^+_\text{d}(\omega)-g^-_\text{d}(\omega)][1-2n_\text{d}(\hbar\omega)],
\end{split}
\label{gf_raK}
\end{equation}
with $n_\text{d}(\epsilon)$ being the QD distribution of the new fermions.

The QD TDOS is defined through the imaginary part of the QD retarded
Green's function,
$\nu_\sigma(\epsilon)\equiv -(1/\hbar\pi)\text{Im}[G_{\text{d}\,\sigma\sigma}^+(\epsilon)]$.

The Keldysh field integral expression for $\nu_\sigma(\epsilon)$ is
\begin{equation}
\begin{split}
&\nu_\sigma(\epsilon)=-\frac{1}{2\pi i\hbar}\frac{1}{\mathcal{N}_0}\underset{\mu\rightarrow\infty}{\text{lim}}e^{\beta\mu}
\int_{-\infty}^\infty dte^{\frac{i}{\hbar}\epsilon t}\times\\
&\!\!\!\times\!\int\mathcal{D}[\chi^\text{cl}(t),\chi^\text{q}(t)]
e^{\frac{i}{\hbar}S_\text{eff}[\chi^\text{cl}(t),\chi^\text{q}(t)]}\times\\
&\!\!\!\times\![\bar{\chi}_-(t)\chi_+(0)G^>_\text{d}(\sigma t|\sigma 0)-\bar{\chi}_+(t)\chi_-(0)G^<_\text{d}(\sigma t|\sigma 0)],
\end{split}
\label{tdos_kfi}
\end{equation}
where $\chi_\pm(t)$ is the slave-bosonic field on the forward and backward
branches of $C_\text{K}$,
\begin{equation}
\begin{split}
&iG^{>,<}_\text{d}(\sigma t|\sigma 0)=\\
&=\biggl[\frac{1}{2}\pm\frac{1}{2}-n_\text{d}(\epsilon_\text{d}+\mu)\biggl]\exp\biggl[-\frac{i}{\hbar}(\epsilon_\text{d}+\mu)t\biggl],
\end{split}
\label{gv_lg}
\end{equation}
As in Refs. \onlinecite{Altland_2010,Altland_2009} the distribution
$n_\text{d}(\epsilon)$ is the double step. In the present context this is due
to the noninteracting nature of the new fermions.

Since the effective Keldysh action
$S_\text{eff}[\chi^\text{cl}(t),\chi^\text{q}(t)]$ is quadratic, the
functional integral in Eq. (\ref{tdos_kfi}) may be performed exactly for any
real $\mu>0$. The limit $\mu\rightarrow\infty$ is readily taken
afterwards. The final nonperturbative result for the QD TDOS is
\begin{equation}
\nu_\sigma(\epsilon)=\frac{\mathcal{P}(\epsilon)}
{[\epsilon_\text{d}-\epsilon+(g/\hbar)\Sigma_\text{R}(\epsilon)]^2+[(g/\hbar)\Sigma_\text{I}(\epsilon)]^2},
\label{tdos_f}
\end{equation}
where $\Sigma_\text{R}(\epsilon)$, $\Sigma_\text{I}(\epsilon)$ are the real
and imaginary parts of the retarded slave-bosonic self-energy for which we
find the following analytical expressions,
\begin{equation}
\begin{split}
&\Sigma_\text{R}(\epsilon)=\hbar[D(\epsilon)/\pi]\{\epsilon/2W+\\
&\!\!\!+\!(1/2\pi)\text{Re}\!\!\sum_x[(1\!+\!i\epsilon/W)\psi[1/2\!-\!i\beta(iW\!\!+\!\mu_x)/2\pi]-\\
&-\psi[1/2-i\beta(\mu_x-\epsilon)/2\pi]]\},\\
&\Sigma_\text{I}(\epsilon)=\hbar[D(\epsilon)/2\pi]\sum_xn_x(\epsilon),
\end{split}
\label{rsbse_r_i}
\end{equation}
where $\psi(x)$ is the digamma function and $n_x(\epsilon)$ are the contacts
Fermi-Dirac distributions, and the numerator is
\begin{equation}
\begin{split}
&\mathcal{P}(\epsilon)=\frac{gD(\epsilon)}{2\pi^2}\frac{n_\text{L}(\epsilon_\text{d})+n_\text{R}(\epsilon_\text{d})-
2n_\text{L}(\epsilon_\text{d})n_\text{R}(\epsilon_\text{d})}{1-n_\text{L}(\epsilon_\text{d})n_\text{R}(\epsilon_\text{d})}.
\end{split}
\label{numer}
\end{equation}
The imaginary part in Eq. (\ref{rsbse_r_i}) has a particularly clear physical
meaning. The slave-boson describes the empty state of the QD. From
Eq. (\ref{rsbse_r_i}) it follows that for energies above the chemical
potentials $\mu_x$ the imaginary part of the slave-bosonic self-energy
$\Sigma_\text{I}(\epsilon)\rightarrow 0$ which means that the life-time of the
empty state goes to infinity, {\it i.e.}, the QD is empty. In contrast, below
$\mu_x$ the imaginary part is finite leading to a finite life-time of the
empty state, {\it i.e.}, the QD is filled.

It is important to emphasize that the QD TDOS, Eq. (\ref{tdos_f}), is finite
at any finite temperature but logarithmically diverges for $T=0$ at the
chemical potentials. Thus, one expects that for very low temperatures the
simple quadratic theory is not valid. This is in accordance with what one
generally expects on purely theoretical grounds \cite{Altland_2010} from
expansions around zero field configurations.  Zero field expansions complement
low temperature theories, being expansions around non-zero field
configurations, like mean field theories
\cite{Aguado_2000,Ratiani_2009}. These non-zero field expansions fail to
describe the high-temperature behavior of the Kondo effect in QDs as we have
demonstrated in Fig. \ref{figure_1}. Therefore, one concludes that for a
comprehensive description of the Kondo effect in the whole temperature range
it is desirable to have in one's disposal both types of field theories.

Let us clarify the nature of the approximation used for the tunneling action,
Eq. (\ref{sbta}), and the applicability range of the result for the TDOS,
Eq. (\ref{tdos_f}). Formally the small parameter of the expansion is the
tunneling matrix element $T$. Hence, $\Gamma=2g/\pi$ should be small. However,
the tunneling matrix elements always enter in product with the slave-bosonic
fields. Thus our theory will be valid in physical situations where the large
values of the slave-bosonic amplitude are not important, {\it i.e.}, when the
probability of the QD empty state is not large. On one side this happens when
$\mu_0-\epsilon_\text{d}$ is large and on the other side when the temperature
is not too low and thus the Kondo resonance is not too strong so that the
fluctuations of the QD empty state are not too large. Therefore, we assume
that our theory should be applicable for
\begin{equation}
\mu_0-\epsilon_\text{d}\gtrsim\Gamma\quad\text{and}\quad T\gtrsim T_\text{K}.
\label{criteria}
\end{equation}
One should keep in mind that this is a crude estimate and our theory may
still work qualitatively and perhaps semiquantitatively outside those
inequalities. As demonstrated in the next section, our simple quadratic
approximation, accounting for the physics in the vicinity of the zero
slave-bosonic field configuration, provides a reasonable description of the
Kondo physics. It also gives a finite single-particle resonance. This situation
is definitely better than the one taking place in perturbative approaches
\cite{Sivan_1996} which have divergences at the single-particle resonance.
Our theory being nonperturbative avoids this problem but requires the next
order term, {\it i.e.}, the one which is quartic in the slave-bosonic field,
in the expansion of the tunneling action (\ref{sbta_ex}) for a better
quantitative description. This will be the subject of a more advanced theory.
Here we only note that this advanced theory may be constructed in terms of the
quadratic theory presented in this work. Namely, the present theory will play
a role of a nonperturbative "free propagator" for the quartic theory.

It is interesting to look at the result for the QD TDOS, Eq. (\ref{tdos_f}),
from the physical point of view. Within the range of its applicability it
suggests that the quasiparticle state in the QD represents a superposition of
the bare electron state and the empty state of the QD coupled to
contacts. Thus in this range the life-time of the QD quasiparticles can be
estimated as the life-time of the QD empty state. This may be very attractive
for the experiments measuring the quasiparticle life-times since observing QD
states is easier than a direct observation of its quasiparticles.

Finally, we would like to emphasize an additional fundamental advantage of our
method: our field-theoretic approach is a truly field theory in contrast with
mean field theories \cite{Aguado_2000,Ratiani_2009}. The point is that here we
work only with a physical field, being the slave-bosonic field, and not with
artificial fields like the Lagrange multipliers fields used in mean field
theories. Thus our theory has more transparent access to physics avoiding such
artefacts of mean field theories as slave-bosonic condensation.
\begin{figure}
\includegraphics[width=7.6 cm]{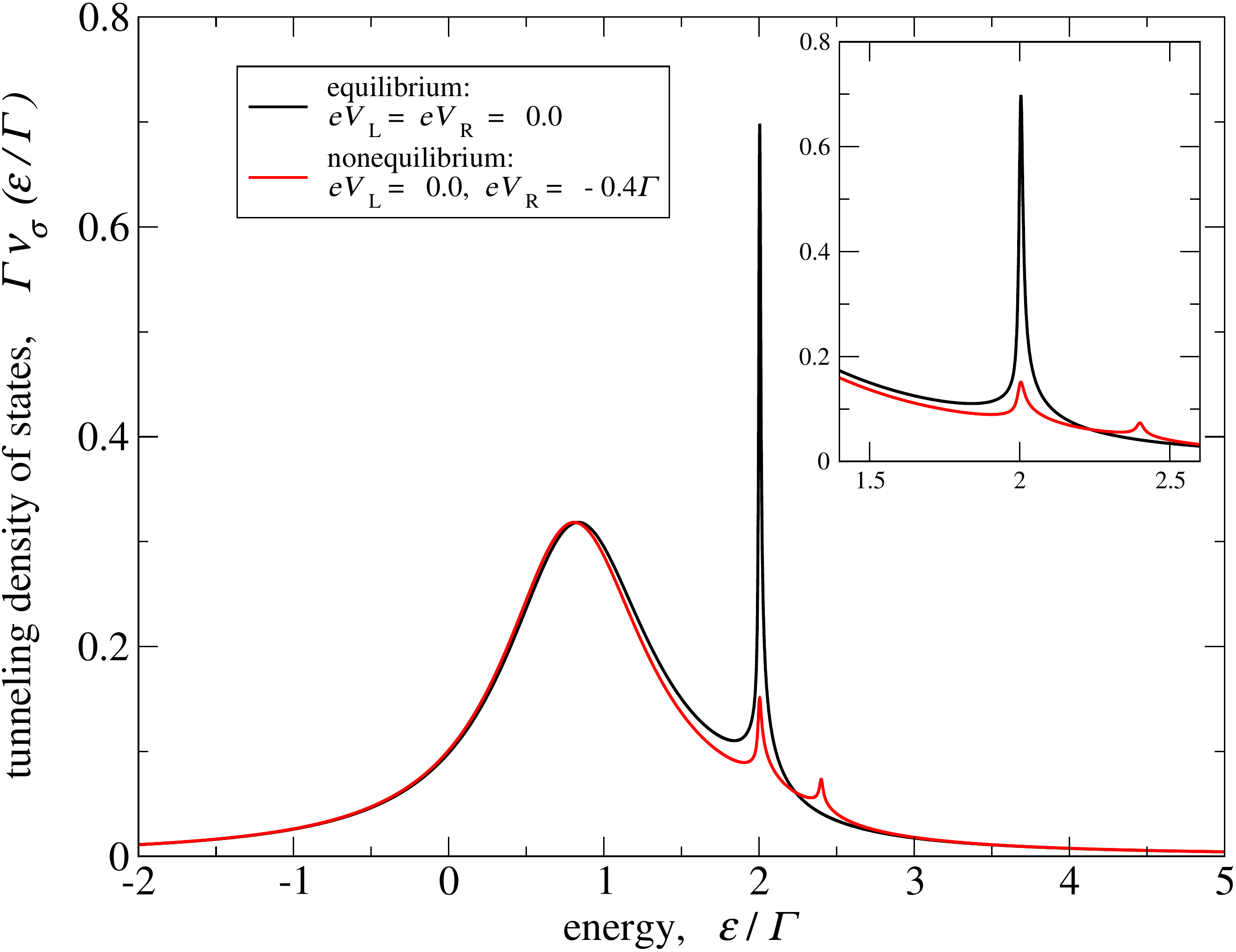}
\caption{\label{figure_2} (Color online) The equilibrium and nonequilibrium
  TDOS in the Kondo regime. Here $kT=0.005\Gamma$,
  $\mu_0-\epsilon_\text{d}=1.876\Gamma$, $W=100\Gamma$. In equilibrium, $V=0$,
  there is a sharp many-particle peak at the Fermi energy. In nonequilibrium,
  $V\neq0$, the peak is reduced and split into two lower peaks (see
  inset). The total spectral weights of the equilibrium and nonequilibrium
  situations are almost the same differing by 1\% which is due to the accuracy
  of the numerical integration involved in the total spectral weight.}
\end{figure}

\section{Discussion of the results}\label{dr}
\begin{figure}
\includegraphics[width=7.6 cm]{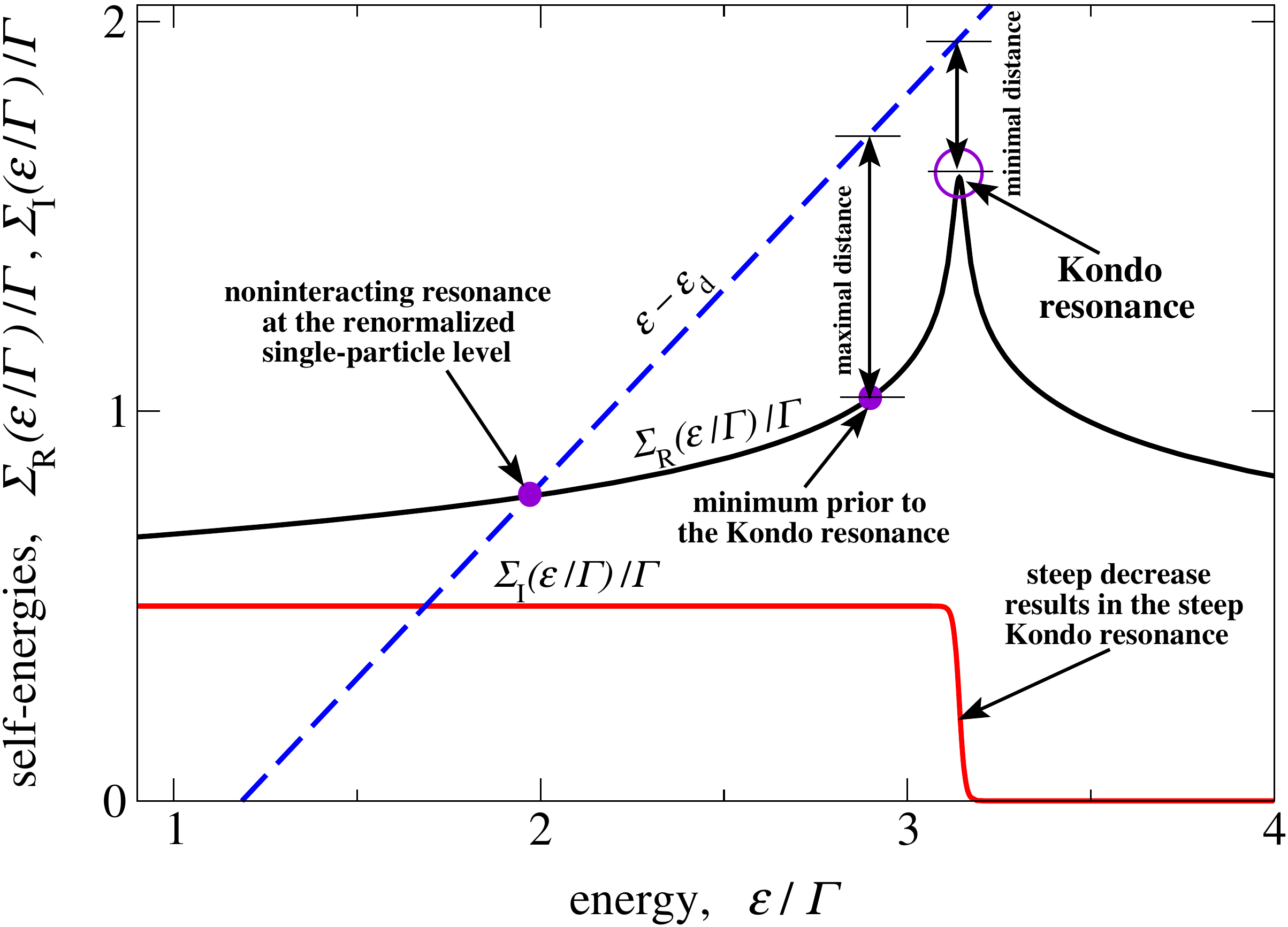}
\caption{\label{figure_3} (Color online) The mechanism of the Kondo resonance
  formation in the nonperturbative Keldysh field theory in the vicinity of the zero
  slave-bosonic field configuration. Here $V=0$, $kT=0.008\Gamma$,
  $\mu_0-\epsilon_\text{d}=1.95\Gamma$.}
\end{figure}

In Fig. \ref{figure_2} the QD TDOS (\ref{tdos_f}) is shown. In addition to the
single-particle resonance at the renormalized QD noninteracting energy level
it reveals a many-particle peak at the Fermi energy, known as the Kondo
resonance. The formation of the Kondo resonance is explained in Fig.
\ref{figure_3} as an interplay between $\Sigma_\text{R}(\epsilon)$ and
$\Sigma_\text{I}(\epsilon)$ from (\ref{rsbse_r_i}). The distance between
$\epsilon-\epsilon_\text{d}$ and $\Sigma_\text{R}(\epsilon)$ reaches a minimum
where $\Sigma_\text{R}(\epsilon)$ has its maximum. This happens at the Fermi
energy. At the same time $\Sigma_\text{I}(\epsilon)$ has a steep decrease at
the Fermi energy. Therefore, the two terms in the denominator of Eq.
(\ref{tdos_f}) become both minimal at the Fermi energy giving rise to the
Kondo resonance.
\begin{figure}
\includegraphics[width=7.6 cm]{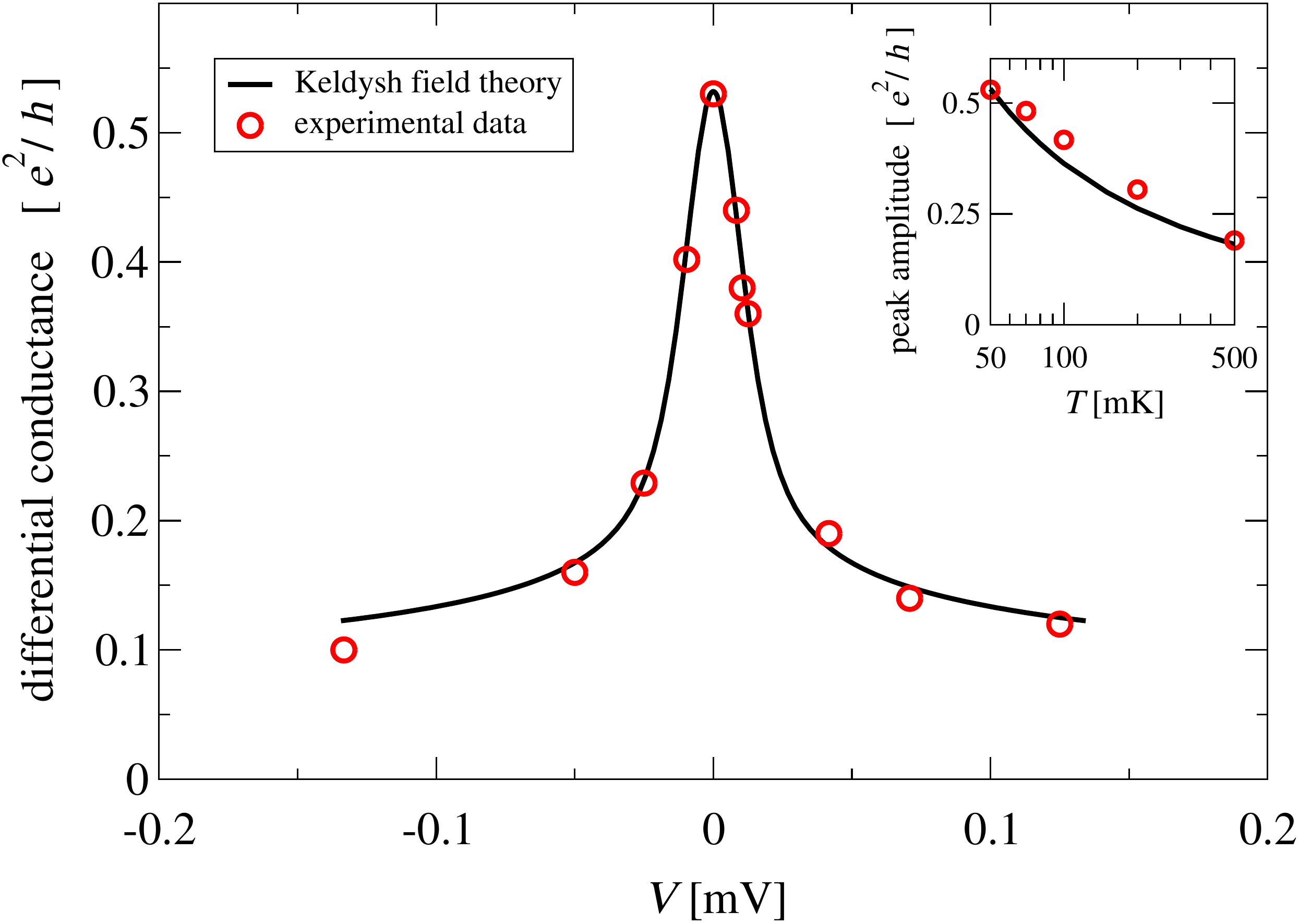}
\caption{\label{figure_4} (Color online) The Kondo peak in the differential
  conductance. Inset shows the temperature dependence of the differential
  conductance maximum at $V=0$. The solid line is the result of our
  nonperturbative Keldysh field theory. The circles show the experimental data
  of Ref. \onlinecite{Ralph_1994}.}
\end{figure}

To verify our field-theoretic description of the Kondo physics we first
compare it with experimental data. The presence of the Kondo resonance in the
QD TDOS has an impact on the other QD observables. In particular, experiments
\cite{Ralph_1994} show a peak in the differential conductance at $V=0$. Using
the expression for the current (see Eq. (3) in Ref. \onlinecite{Wingreen_1994})
through a QD together with Eqs. (\ref{tdos_f}) and (\ref{rsbse_r_i}) we obtain
this behavior of the differential conductance shown in Fig. \ref{figure_4}. To
get Fig. \ref{figure_4} we have taken the values of the parameters,
$\Gamma=2.6875$ meV, $W=5$ eV, $T=50$ mK,
$\Gamma/[\pi(\mu_0-\epsilon_\text{d})]=0.1224$, close to the ones which were
estimated in Ref. \onlinecite{Ralph_1994}.
\begin{figure}
\includegraphics[width=7.6 cm]{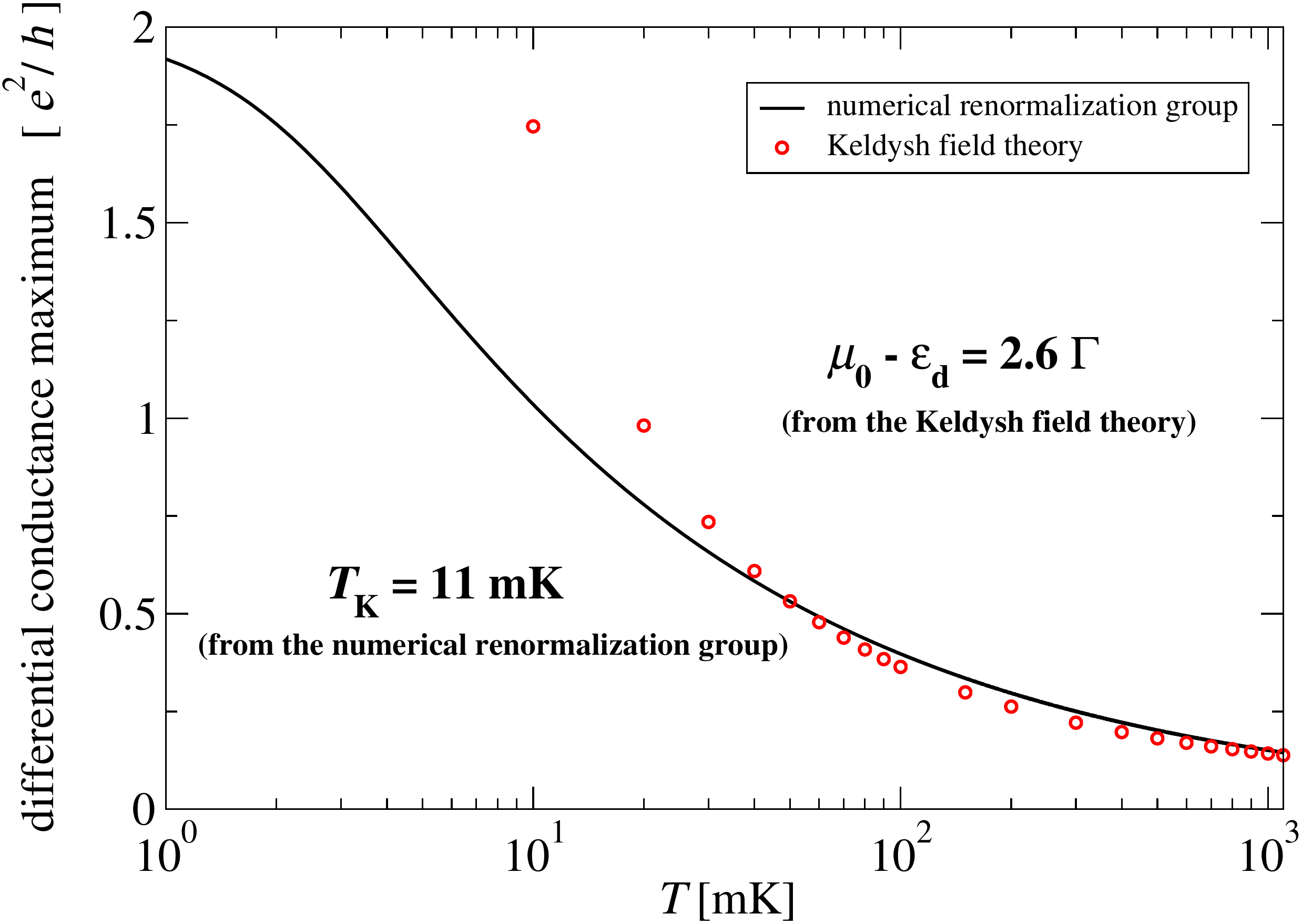}
\caption{\label{figure_5} (Color online) Comparison of our field theory with
  the numerical renormalization group theory. The solid line is obtained from
  the numerical renormalization group calculations
  \cite{Costi_1994,Goldhaber-Gordon_1998a,Grobis_2008}. The circles show the
  differential conductance maximum obtained from our field theory for the
  values of the parameters used in Fig. \ref{figure_4}.}
\end{figure}
\begin{figure}
\includegraphics[width=7.6 cm]{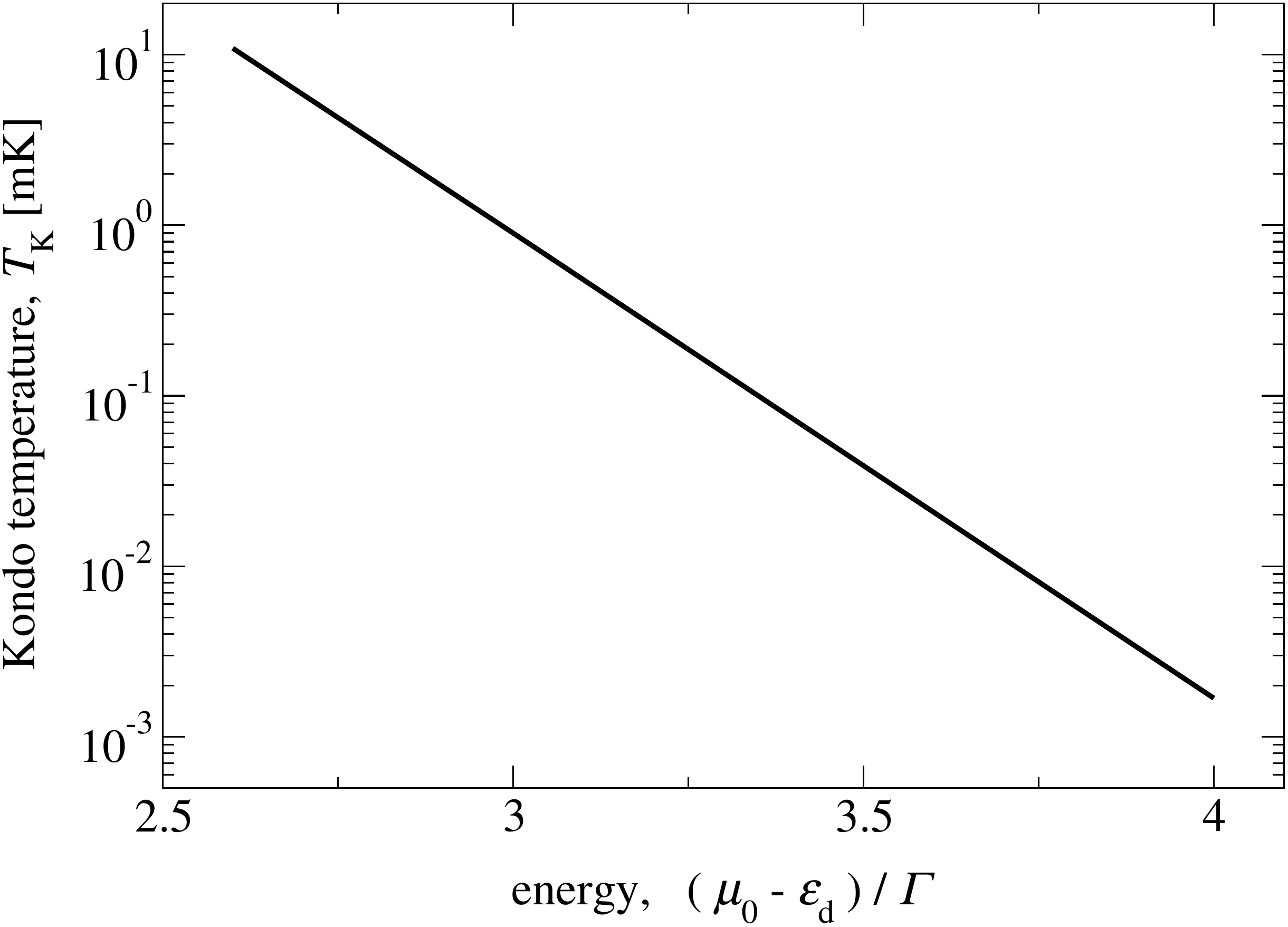}
\caption{\label{figure_6} (Color online) Keldysh field-theoretic prediction
  of the Kondo temperature dependence on the QD single-particle energy level
  $(\mu_0-\epsilon_\text{d})/\Gamma$.}
\end{figure}

\begin{figure}
\includegraphics[width=7.6 cm]{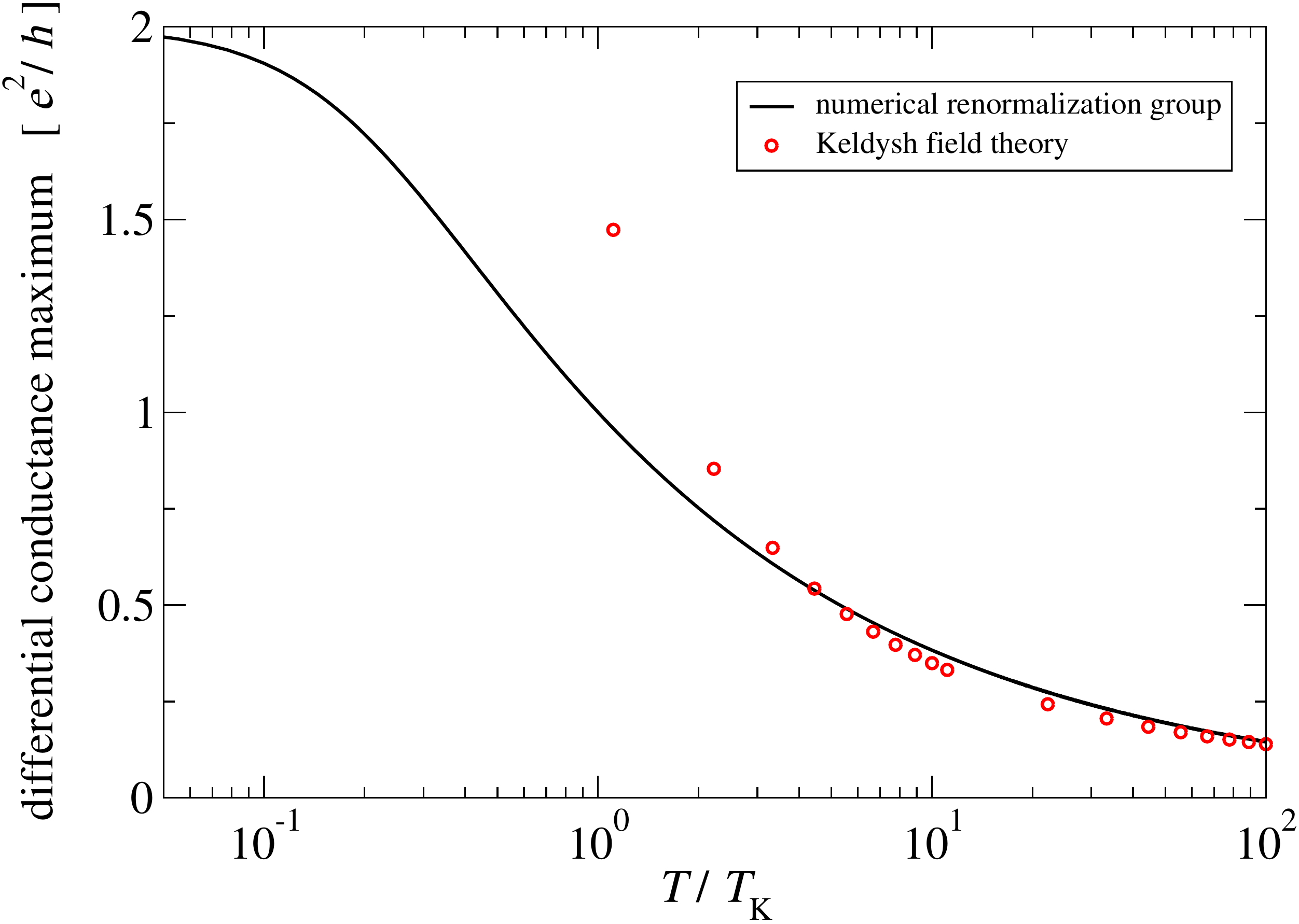}
\caption{\label{figure_7} (Color online) Comparison of the universal
  temperature dependence of the differential conductance maximum in our field
  theory and in the numerical renormalization group theory. The solid line is
  the result of the numerical renormalization group calculations
  \cite{Costi_1994,Goldhaber-Gordon_1998a,Grobis_2008}. The circles show the
  result obtained from our field theory.}
\end{figure}

We would like further to compare our Keldysh field theory with existing
theoretical approaches, in particular, with the numerical renormalization
group theory from Ref. \onlinecite{Costi_1994} which was successfully employed
to describe experiments \cite{Goldhaber-Gordon_1998a,Grobis_2008} on the Kondo
effect in QDs. As we have argued that our theory must be valid for
temperatures $T\gtrsim T_\text{K}$, within this temperature range the
differential conductance maximum in our theory must have the same temperature
dependence as the one in Refs. \onlinecite{Goldhaber-Gordon_1998a,Grobis_2008}
(see, {\it e.g.}, Eq. (2) in Ref. \onlinecite{Goldhaber-Gordon_1998a}, where
we take $s=0.21$) for the case of a symmetric coupling. From this
high-temperature comparison, for each value of $\mu_0-\epsilon_\text{d}$ used
to calculate the differential conductance maximum in the Keldysh field theory,
we can fix the value of $T_\text{K}$ to be used in the empirical form of Ref.
\onlinecite{Goldhaber-Gordon_1998a}. For example, for the parameters used in Fig.
\ref{figure_4} we get the temperature dependence of the differential conductance
maximum shown in Fig. \ref{figure_5}. The Kondo temperature $T_\text{K}=11$ mK agrees
with the rough estimate $T_\text{K}<50$ mK given in Ref. \onlinecite{Ralph_1994},
where a weak asymmetry in the capacitance of the QD to the left and right contacts
was assumed. The same comparison between the Keldysh field theory and the numerical
renormalization group theory can be done for any value of $\mu_0-\epsilon_\text{d}$
for which the Keldysh field theory is applicable. In this way we determine how in
our Keldysh field theory $T_\text{K}$ depends on $\mu_0-\epsilon_\text{d}$. This
dependence is shown in Fig. \ref{figure_6} and it is in full accordance with the
standard expression \cite{Hewson_1997},
\begin{equation}
\frac{T_\text{K}}{\Gamma}\thicksim\exp\biggl[-2\pi\frac{\mu_0-\epsilon_\text{d}}{\Gamma}\biggl],
\label{TK}
\end{equation}
where our definition of $\Gamma$ is twice that of Ref. \onlinecite{Meir_1993} (see
also the caption of Fig. \ref{figure_1}). This proves that our theory for the
Kondo effect in QDs correctly predicts the Kondo temperature. Moreover, our theory,
within its applicability range, also predicts that the differential conductance
maximum has a universal temperature dependence with the scaling given by the Kondo
temperature, Eq. (\ref{TK}). This universal temperature dependence is shown in
Fig. \ref{figure_7} and additionally proves that our Keldysh field theory,
within its applicability range, correctly describes the Kondo physics in
QDs. As one can see from Fig. \ref{figure_7} the Keldysh field-theoretic
description is quantitatively reliable for temperatures
$T\geqslant 2T_\text{K}$, which perfectly agrees with our theoretical
prediction made above in Section \ref{fts}.

Finally, we would like to say a few words about the numerical consistency of
our theory. To do this, we employ the sum rule given by Eq. (39) of
Ref. \onlinecite{Wingreen_1994}. In NCA this sum rule is always satisfied
within 0.5\%. In our theory this depends on how well the applicability
criteria, Eq. (\ref{criteria}), of the Keldysh field theory are satisfied. For
example, for the parameters presented in Fig. \ref{figure_2} the sum rule is
satisfied within 15\% while for $\mu_0-\epsilon_\text{d}=2.5\Gamma$ it is 8.5\%
and for $\mu_0-\epsilon_\text{d}=4.0\Gamma$ it is 4.4\%. One should note that
the sum rule is an integral estimate over the whole energy range. Thus, the
error is gained over the whole range of energies. At the same time for a given
energy the QD TDOS may have higher accuracy. Since the only approximation was
the truncation of all the terms of higher orders than the terms quadratic in
the slave-bosonic field, to improve the consistency of the method and extend
its applicability criteria one should go beyond the quadratic approximation
and this will be done in a subsequent study.

\section{Conclusion}\label{concl}
We have developed a basic slave-boson nonperturbative Keldysh field theory for
the Kondo effect in quantum dots. The theory deals with the physics in the
vicinity of the zero slave-bosonic field configuration where, as we have
shown, the main fraction of the Kondo physics is located at experimentally
relevant temperatures. The presented theory has a closed analytical solution
for the quantum dot tunneling density of states and, despite being relatively
simple, properly describes experimental data on the Kondo peak observed in the
differential conductance, correctly predicts the Kondo temperature and, within
its applicability range, has the same universal temperature dependence of the
conductance as the one obtained in numerical renormalization group calculations.
Therefore, it represents a convenient basis, as a free nonperturbative propagator,
for more advanced theories which could extend the applicability of our approach
to larger values of the slave-bosonic amplitude and, thus, to temperatures much
lower than the Kondo temperature.

\section{Acknowledgments}
The authors thank Alexander Altland and Dmitry Ryndyk for fruitful
discussions. Support from the DFG under the program SFB 689 is acknowledged.

\end{document}